\documentclass{revtex4}

\usepackage{graphicx}
\begin{document}
\bibliographystyle{revtex}

\def\ba{\begin{eqnarray}}
\def\ea{\end{eqnarray}}
\def\bq{\begin{equation}}
\def\eq{\end{equation}}
\def\lsim{\mathrel{\raisebox{-.6ex}{$\stackrel{\textstyle<}{\sim}$}}}
\def\gsim{\mathrel{\raisebox{-.6ex}{$\stackrel{\textstyle>}{\sim}$}}}
\newcommand{\sla}[1]{/\!\!\!#1}

\preprint{CALT-68-2356}
\preprint{MADPH-01-1246}

\title{VLHC predictions for $H\to\tau\tau$ in weak boson fusion}



\author{T.~Plehn}
\affiliation{Department of Physics, University of Wisconsin, 
Madison, WI 53706, USA}

\author{S.~Su}
\affiliation{California Institute of Technology, Pasadena, CA 91125, USA}

\author{D.~Zeppenfeld}
\affiliation{Department of Physics, University of Wisconsin, 
Madison, WI 53706, USA}



\begin{abstract}
Higgs production in weak boson fusion with subsequent decay $H\to \tau\tau
\to e\mu\sla p_T$ provides a means to measure Higgs Yukawa couplings and 
Higgs interactions to weak bosons. The potential precision of cross section
measurements at a VLHC is investigated.
\end{abstract}


\maketitle



Precision measurements of Higgs boson properties are an important goal
of any post-LHC accelerator. The LHC is expected to yield a determination 
of the Higgs boson mass at the $10^{-3}$ level~\cite{ATLAS,CMS}. A 
combination of observations of the Higgs boson, in various decay channels
($H\to\gamma\gamma,\;WW,\;ZZ,\;\tau\tau$), in inclusive searches and 
in the weak boson fusion (WBF) channel, is expected
to yield measurements of various Higgs partial widths with accuracies of
order 10--20\%~\cite{zknr}. A linear collider (LC) can improve these 
measurements by up to one order of magnitude, achieving
a clean separation of individual couplings~\cite{lc}.

In contrast, very little is known about the capabilities of a higher energy
hadron collider for improving our knowledge of the Higgs sector. In this note
we analyze a particular weak boson fusion process, $qq\to qqH,\; H\to\tau\tau$
with subsequent leptonic tau decays, within the SM, 
as a case study for possible VLHC 
reach in precision measurements of Higgs couplings. Unlike gluon fusion, WBF
has small NLO QCD corrections, thus promising small systematic uncertainties. 
Second, the $H\to\tau\tau$ decay modes~\cite{prz_tautau} 
are key processes in the coupling determination at the LHC.
We closely follow this previous LHC analysis and determine signal rates
and cross sections for the main physics backgrounds in $pp$ collisions
at $E_{\rm cm}=50$, 100, and 200 TeV. These cross sections allow 
for a first estimate of the statistical accuracy of coupling measurements at a 
VLHC.


The signal process is $qq\to qqH$, which is mediated by $t$-channel $W$ and
$Z$ exchange. We consider $H\to\tau\tau$ decays with subsequent leptonic 
decays of both taus, i.e. the final state is $jj\ell^+{\ell'}^-\sla p_T$
where, for simplicity, we only consider the $e^\pm\mu^\mp$ 
combinations.
In a previous LHC 
analysis~\cite{prz_tautau} it was found that the physics backgrounds,
$Zjj$ production with $Z\to\tau^+\tau^-$, dominate overall backgrounds.
Reducible backgrounds are mostly from leptonic decays of $W$ pairs, e.g. 
from top-decays or $WWjj$ events. These can be distinguished from tau pairs
by the angular correlation of charged lepton momenta and 
the missing momentum vector.
Thus, we only consider QCD $Zjj$ and electroweak $Zjj$ production, i.e.
$\tau^+\tau^-jj$ production at order $\alpha_s^2\alpha^2$ and $\alpha^4$,
respectively. These backgrounds and the signal are generated as in 
Ref.~\cite{prz_tautau} and include off-shell $Z$'s and photon-$Z$ 
interference.

As compared to the LHC analysis we impose a higher $p_T$ of the 
two forward ``tagging jets'' and a larger dijet invariant mass of these
two candidate quark jets. Specifically the signal is required
to have two jets and two charged leptons with 
\ba
\label{eq:basic}
& p_{T_j} \geq 30~{\rm GeV} \, , \; \; |\eta_j| \leq 5.0 \, , \; \;
\triangle R_{jj} \geq 0.6 \, , \nonumber\\
& p_{T\ell_1} \geq 20~{\rm GeV} \, ,\; \;
p_{T\ell_2} \geq 10~{\rm GeV} \, ,\; \;
|\eta_{\ell}| \leq 3 \, , \; \; \triangle R_{j\ell} \geq 0.6 \, ,
\ea
the two jets must be separated by at least 4.2 units of pseudorapidity,
reside in opposite detector hemispheres and the two charged leptons 
must lie between the two jet definition cones of radius 0.6. The 
neutrinos must lead to missing transverse momentum of 50~GeV or more
and we require
\begin{equation}
\label{eq:mjj}
M_{jj} > 1000~{\rm GeV}
\end{equation}
for the invariant mass of the two tagging jets. All other cuts are as in
Ref.~\cite{prz_tautau}. 

The surviving $e\mu\sla p_T+2$~jet events allow a reconstruction of the 
original tau-pair invariant mass by making use of the large boost of the 
individual taus and their decay products~\cite{tautaumass}: the charged
leptons trace the original tau directions and the transverse momenta of $e$,
$\mu$ and missing neutrinos allow to solve for the $\tau^+$ and $\tau^-$
energies. The reconstruction relies on good missing transverse momentum 
resolution. We assume a performance of 
VLHC detectors equivalent to ATLAS, i.e. energy smearing and $\sla p_T$ 
resolution is handled as in Ref.~\cite{prz_tautau} and reproduces ATLAS
simulations for $H/A\to\tau\tau$ events~\cite{cavalli}.

 \begin{figure}
 \includegraphics{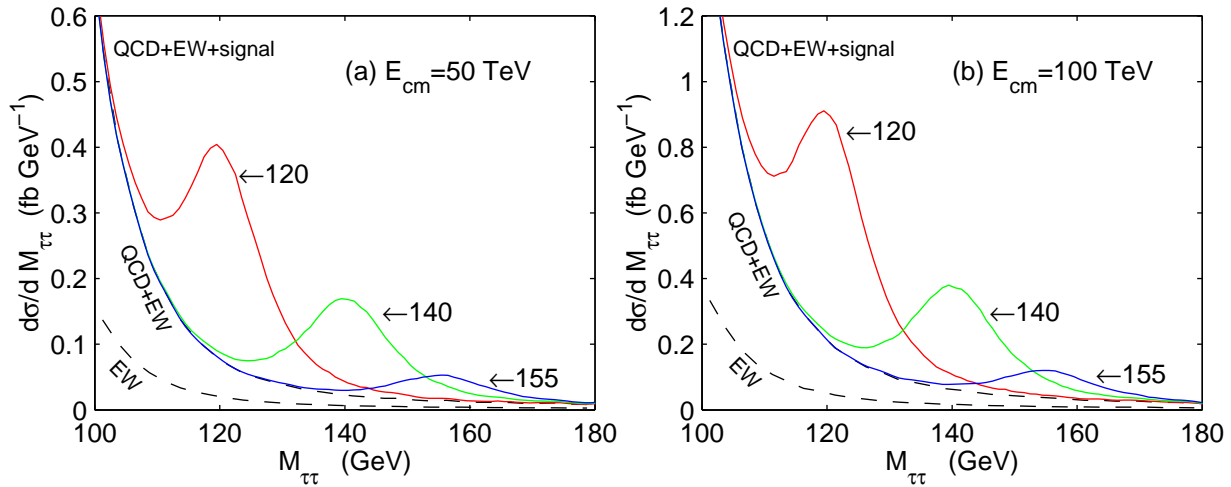}%
 \caption{Reconstructed tau-pair invariant mass distribution at a VLHC of
 (a) 50 TeV and (b) 100 TeV center of mass energy. Signal peaks for 
 $m_H=120$, 140, and 155~GeV are added to the combined physics backgrounds 
 from electroweak and QCD $\tau\tau jj$ production. See text for cuts.}
 \label{fig:tautaumass}
 \end{figure}

The resulting reconstructed $M_{\tau\tau}$-spectra are shown in 
Figure~\ref{fig:tautaumass} for VLHC energies of 50 and 100~TeV.
The Higgs peak is clearly visible above the physics backgrounds,
even for Higgs masses as large as 155~GeV. The statistical significance 
of the signal can be estimated by considering the signal and background rates 
in a reasonable mass window, which we take as 
$m_H-15~{\rm GeV}<M_{\tau\tau}<m_H+15~{\rm GeV}$.

Resulting cross sections are given in Table~\ref{tab:rates}.
It is quite remarkable that we may expect signal to background ratios well
above 1 over the entire intermediate mass range. This is combined with large
signal rates, of order hundreds to a thousand, for an integrated luminosity
of 100~fb$^{-1}$. Given the expected number of signal events, $S$, and
background events, $B$, we can estimate the statistical error with which the 
signal cross section can be estimated as 
$\Delta\sigma_H/ \sigma_H = \sqrt{S+B}/S$.
The corresponding statistical errors are listed in the last column of 
Table~\ref{tab:rates} and reside in the 5--10\% range for a 50~TeV machine
and Higgs masses between 120 and 150 GeV, and improve to the 3--6\% range
for a 200 TeV machine.

The numbers presented in the left half of 
Table~\ref{tab:rates} do not include  efficiency 
factors for jets or for lepton identification and isolation. In previous 
studies these factors amounted to overall efficiency factors of 0.67 for 
signal and backgrounds~\cite{prz_tautau}. 
(Note that geometric acceptance is included in our
simulation.) Other reduction factors for the signal will arise from parton 
shower corrections: including QCD radiation, the selection of the 2 quark jets
is not unique. Misidentified tagging jets then tend to fall below 
the $M_{jj}$ threshold of Eq.~(\ref{eq:mjj}). Since QCD radiation is 
relatively suppressed for the $t$-channel color singlet exchange of the 
signal, this effect is modest~\cite{wbfexp} and we may assume an overall 
efficiency of $\epsilon_{\rm WBF}=0.5$ for WBF processes compared to 
$\epsilon=0.67$ for the QCD $Zjj$ background. 

The cross sections of Table~\ref{tab:rates} do not yet make use of another
characteristic of WBF events: they are mediated by the $t$-channel exchange of
a colorless object. Destructive interference between initial and final state 
radiation in such processes leads to suppressed gluon radiation in the 
central region, between the two quark jets~\cite{bjgap}. QCD 
background processes, which are dominated by $t$-channel gluon exchange, 
predominantly radiate in the central region. This difference can be exploited
by a veto on central jets, in the region
\bq
\eta_{j,min}^{\rm tag} +0.6  <  \eta_j^{\rm veto}
< \eta_{j,max}^{\rm tag} -0.6\; . \label{eq:etaveto}
\eq
A detailed study on feasible $p_T$ thresholds at a VLHC is still needed. For 
the LHC, an parton level analysis indicates that a veto of central
jets of $p_T>20$~GeV will suppress the signal by a factor $P_{\rm surv}=0.9$,
while analogous factors for the EW and the QCD $Zjj$ backgrounds are 0.75
and 0.3, respectively~\cite{DR_thesis}. Due to the increased $M_{jj}$ 
values and the larger rapidity distance between the quark jets of the signal,
these numbers may improve at a VLHC. However, higher $p_T$ thresholds
of vetoed jets will
counteract such improvements. We take the LHC values as an educated guess
for a VLHC as well. Including these efficiency factors and survival 
probabilities, signal and background rates are given in the right half of 
Table~\ref{tab:rates}. Signal to background rates improve further, due to the 
background suppression of the central jet veto. However, the statistical 
accuracy of the cross section determination suffers somewhat, due to the 
more realistic assumptions on reconstruction efficiencies. Nevertheless,
statistical errors of 5\% or better are possible at a VLHC with a mere
100~fb$^{-1}$ of data. This can be improved significantly by including 
other tau decay channels like hadronic decays of one tau and leptonic 
decays into $e^+e^-\sla p_T,\; \mu^+\mu^-\sla p_T$~\cite{prz_tautau,zknr}. 

A point of concern are Higgs backgrounds. $H\to WW\to e\mu\sla p_T$
events become the dominant reducible $WW$ background for Higgs masses above
125~GeV at the LHC, and contribute about a third of the overall background
at $m_H=150$~GeV~\cite{prz_tautau}.
The same should be expected at a VLHC. In addition,
gluon fusion becomes a more pronounced source of $Hjj$ events at the 
higher VLHC energies~\cite{oleari}, and thus the clean separation of 
gluon fusion and WBF cross sections will require additional efforts. 
However, we do not expect these complications to significantly alter the 
previous optimistic findings.

We conclude that a VLHC can study $H\to\tau\tau$ decays with
high precision in WBF events. Integrated luminosities of  a few hundred 
fb$^{-1}$ provide high statistics and clean samples which allow to measure
$H\tau\tau$ Yukawa couplings with statistical errors at the $10^{-2}$ level.
At the same time the small QCD NLO corrections to WBF promise small systematic
errors. Detailed studies of WBF channels at a VLHC are required to assess
experimental systematic errors.

\begin{table}[t]
\caption{Signal and background cross sections (in fb) for 
$qq\to qqH,\;H\to\tau\tau \to e\mu\sla p_T$ events at a VLHC of center of mass
energy 50, 100, and 200~TeV, for SM Higgs masses between 120 and 160~GeV. 
The $S/B$ and $\sqrt{S+B}/S$ columns give signal to background ratios and
statistical errors for a determination of the signal cross section with 
100~fb$^{-1}$ of data. The first two columns give these estimates without
taking into account reconstruction efficiencies or a central jet veto.
The second set uses LHC inspired assumptions for $\epsilon$ and $P_{\rm surv}$.
See text for details. }
\label{tab:rates}
\begin{tabular}{c|c|cccc|cc|cc}\hline
$E_{\rm cm} $&$m_H$ &Signal 
&QCD&
EW &
${\rm QCD}+{\rm EW}$&
$S/B$&$\sqrt{S+B}/S$ & $S/B$&$\sqrt{S+B}/S$ \\
&&& $\tau\tau{jj}$ & $\tau\tau{jj}$ & $\tau\tau{jj}$ 
& \multicolumn{2}{c|}{$L=100\ {\rm fb}^{-1}$, $\epsilon \equiv 1$}
& \multicolumn{2}{c}{$L=100\ {\rm fb}^{-1}$, $\epsilon = \epsilon_{\rm LHC}$}
\\ \hline
&120&4.98&2.48&0.83&3.31&1.50&5.8\%&2.76&7.8\% \\
&130&3.86&1.04&0.38&1.42&2.72&6.0\%&4.95&8.3\% \\
50 &140&2.48&0.56&0.22&0.78&3.17&7.3\%&5.70&10.3\% \\
TeV&150&1.23&0.36&0.16&0.52&2.38&10.7\%&4.22&14.9\% \\
&155&0.70&0.31&0.14&0.44&1.59&15.3\%&2.81&20.7\% \\
&160&0.23&0.27&0.12&0.39&0.59&34\%&1.04&44\%\\ \hline

&120&10.67&6.84&2.07&8.91&1.20&4.1\%&2.23&5.5\% \\
&130&8.32&2.94&0.95&3.89&2.14&4.2\%&3.96&5.8\% \\
100 &140&5.37&1.57&0.55&2.12&2.53&5.1\%&4.63&7.1\% \\
TeV&150&2.66&1.00&0.38&1.38&1.93&7.6\%&3.49&10.4\% \\
&155&1.51&0.84&0.33&1.17&1.29&10.8\%&2.33&14.5\%\\
&160&0.49&0.71&0.29&1.00&0.49&25\%&0.87&31\%  \\ \hline

&120&20.6&16.9&4.43&21.3&0.96&3.1\%&1.83&4.1\% \\
&130&16.1&7.35&2.04&9.38&1.71&3.1\%&3.23&4.3\% \\
200 &140&10.4&3.98&1.19&5.17&2.01&3.8\%&3.75&5.2\% \\
TeV &150&5.16&2.55&0.81&3.36&1.54&5.7\%&2.84&7.6\% \\
&155&2.92&2.14&0.70&2.84&1.03&8.2\%&1.90&10.8\% \\
&160&0.95&1.83&0.62&2.45&0.39&19.4\%&0.71&23.7\% \\ \hline

\end{tabular}
\end{table}

\vspace{-0.25in}
\begin{acknowledgments}
This research was supported in part by the University of Wisconsin Research
Committee with funds granted by the Wisconsin Alumni Research Foundation and
in part by the U.~S.~Department of Energy under grant
No.~DE-FG02-95ER40896. S.S. was supported by DOE grant DE-FG03-92-ER-40701.
\end{acknowledgments}

\end{document}